\documentclass{aa}  
\usepackage{graphicx}
\usepackage{txfonts}
\usepackage{natbib}
\begin{document}
   \title{Evidence for variable outflows \\ in the  Young
    Stellar Object V645 Cygni}

   \author{A.J. Clarke\inst{1}, S.L. Lumsden\inst{1},
   R.D. Oudmaijer\inst{1}, A.L. Busfield\inst{1}, M.G. Hoare\inst{1},
   T.J.T. Moore\inst{2}, T.L. Sheret\inst{1} \and J.S. Urquhart\inst{1}}

   \offprints{A.J.Clarke}

  \institute{School of Physics and Astronomy, University of Leeds, Leeds, LS2 9JT, UK\\
  \email{ajc@ast.leeds.ac.uk}\and Astrophysics Research Institute,
  Liverpool John Moores University, Twelve Quays House, Egerton Wharf,
  Birkenhead, CH41 1LD, UK  }

  \date{Accepted June 22, 2006}

  \abstract{} {As part of the Red MSX Source Survey of Massive Young
  Stellar Objects (MYSOs) we have conducted multi-wavelength follow up
  observations of the well-known object.  We present our data on this
  object, whose near-infrared spectrum is exceptional and place these
  in context with previous observations} {Our observations of V645 Cyg
  included near/mid infrared imaging observations, $^{13}$CO 2-1 line
  observations and high signal-to-noise velocity resolved
  near-infrared spectroscopy.}  {The spectrum shows P-Cygni hydrogen
  Brackett emission, consistent with a high velocity stellar wind.  A
  red-shifted emission component to a number of near-IR emission lines
  was also uncovered. This is associated with a similar component in
  the H$\alpha$ line. V645 Cyg is also found to have variable CO first
  overtone bandhead emission}{The data clearly indicate that the
  outflow of V645 Cyg is variable. The unidentified feature in a
  previously published optical spectrum is identified with a receding
  outflow at 2000 kms$^{-1}$. The nature of this feature, which is
  found in hydrogen and helium atomic lines and CO molecular lines
  remains a puzzle.  }

  \titlerunning{RMS observations of V645 Cyg}
  \authorrunning{Clarke et al.}  

  \keywords{V645 Cyg; early-type; spectroscopy; circumstellar matter;
  winds,outflows}

   \maketitle
%

\section{Introduction}

The variable star V645 Cyg has been subject to detailed study since
its identification with the strong infrared source AFGL 2789 by
\cite{walker} and \cite{lebofsky}. The optical follow up observations
of \citet{cohen} revealed the star to be a young O7e type star within
a polarised optical nebula. The presence of P-Cygni profile hydrogen
recombination lines in the optical spectrum was interpreted as
evidence for a stellar wind with velocities as high as
1000~\rm{kms}${^{-1}}$. Further analysis by \cite{goodrich} and
\cite{hamann} showed the star to have similar P-Cygni profile He {\sc
i}, O {\sc i}, Si {\sc ii} and Ca {\sc ii} transition lines and concluded
that the optical spectrum was more typical of a Herbig Ae/Be type
star. In addition to the lines associated with the stellar wind, the
optical spectrum of V645 Cyg also possesses emission lines of C {\sc i},
[S{ \sc ii}], K {\sc i}, Fe{\sc i}, Fe {\sc ii} and [Fe {\sc ii}]. The
presence of blue-shifted forbidden [S {\sc ii}] lines had been
suggested by \citet{hamann} to be evidence for a dense circumstellar disc
which obscures the receding, red-shifted part of a bipolar flow around
V645 Cyg.

Near-IR spectroscopic observations have also been performed on V645
Cyg by \cite{harvey}, \cite{geballe}, \cite{carr} and \cite{biscaya}
who found H {\sc i} emission and  that the CO first-overtone
bandheads may be variable respectively.

V645 Cyg is located within a high density molecular cloud and has been
associated with a low velocity moderately collimated bipolar molecular
outflow. The CO millimetre line observations of \cite{rodriguez} and
\cite{montenegro} showed the outflow to be oriented in the north-south
direction at small scales ($\approx$15 arcsec), coinciding with the
optical outflow observed by \cite{goodrich}. Whilst at larger scales
($\approx$1$\arcmin$) the CO outflow is reoriented into the
southeast-northwest direction.

Studies of the radio continuum emission found weak ($\approx$0.8 mJy)
extended ($\approx$10 arcsec) emission in the north-south direction
(\citealt{curiel}; \citealt{skinner}). This emission was also determined to
be variable by \cite{girart}, who speculated that the variability
could be due to the episodic ejection of material or recombination of
the ionised gas on a very short timescale.  The object is also
associated with OH (1665 MHz) \citep{morris}, H$_2$O \citep{lada} and
methanol masers \citep{slysh}.

Therefore, the picture of V645 Cyg that has emerged is that of a
relatively unembedded young massive star, with a high velocity stellar
wind and an associated optical and molecular outflow.

We are currently undertaking a survey aimed at finding massive young
stellar objects (MYSOs) in the Galaxy (\citealt{lumsden};
\citealt{hoare}), known as the Red MSX Survey (RMS). We have selected
the MSX source G094.6028-01.7966 as having infrared colours indicative
of a MYSO. This source is identified as V645 Cygni. In order to have
complete datasets to study the properties of MYSOs, we conducted a
series of follow-up observations.  Here, we present millimetre,
mid-IR, near-IR and optical observations of V645 Cyg. In Section
\ref{observations} we describe the observations and the procedure
taken to produce the final spectrum, in Section \ref{results} we
present the data and discuss the features present and in Section
\ref{discuss} we discuss our observations in the context of previous
work.


\section{Observations and Data Reduction}
\label{observations}

\subsection{Near-IR imaging}

The near-IR imaging observations were obtained using the United
Kingdom Infrared Telescope (UKIRT) Fast Track Imaging (UFTI) camera on
the 5th of August 2002. UFTI has a 1024$\times$1024 HgCdTe array with
a 0.091 arcsec per pixel plate scale. The $K$ band filter with 2.03
to 2.37$\mu$m coverage was used for the observations. To correct for
sky background noise the observations were performed using a standard
9 point jitter pattern. The bias, flat and dark correction together
with mosaicing was performed using the ORAC-DR UFTI
package\footnote{http://www.oracdr.org/}. The ORAC-DR pipeline
generated final mosaic image was corrected for astrometry by using Two
Micron All Sky Survey (2MASS) field sources as references.

\subsection{Mid-IR imaging}

The mid-IR imaging data was obtained using the Mid-Infrared Echelle
Spectrometer (Michelle) on UKIRT on the 2nd of August 2002. Michelle
is a mid-infrared (8-25 micron) imager and spectrometer with a 320x240
Si:As array. In imaging mode Michelle provides a 67.2 x 50.4 arcsec
field of view with a 0.2134$\arcsec$ per pixel plate scale. The
observations were made using a narrow $N$ band (12.5$\mu$m) filter
with a bandwidth of 1.2$\mu$m. As is standard at these wavelengths,
the imaging was performed using an on chip nod and chop procedure. The
nodding was performed in Right Ascension with a period of 20.5 sec and
the chopping in Declination with a frequency of 6.34 Hz. The chop
throw was 20.7 arcsec and the nod separation was 20 arcsec.  The
airmass during the observations was 1.57. The data were reduced using
the ORAC-DR Michelle\footnotemark[1] package and the nod and chop
pairs were median averaged to produce a final 20 arcsec $\times$ 20
arcsec image.  The resulting near- and mid-infrared images are shown
in Fig.~\ref{morphology}. Note that the central source visible in the
{\it K} band image, known as N0 is also present in the mid-IR, whereas
the N1 component, located to the North-West is not. The astrometry is
derived from the data by comparison with standard stars. The mid-IR
position of the source is offset to the East by less than 2 arcsec
compared to the 2MASS position of N0, and is within UKIRT's pointing
accuracy.

\subsection{Near-IR spectroscopy}

The spectroscopy data was obtained using the UKIRT 1-5$\mu$m Imager
Spectrometer (UIST) on the night of 12 June 2003. UIST has a
1024$\times$1024 InSb Array and a 0.12 arcsec per pixel plate
scale. The observations were made using a 120 arcsec long slit with a
width of 4 pixels. The seeing was stable at 1 arcsec throughout the
observations. The HK grism was used, which allowed complete wavelength
coverage from 1.4 $\mu$m to 2.5 $\mu$m with an approximate spectral
resolution of $ \lambda / \delta \lambda \approx$ 450. This
corresponds to approximately 650 \rm{kms}$^{-1}$ at 1.6 $\mu$m. The
slit was aligned to 120$^{\rm o}$ East of North.  The observations
were nodded along the slit to provide sky subtraction and were
preceded by a flat and argon arc exposure for suitable later
calibration.

A standard star was observed immediately before the object so that
telluric absorption could be corrected for, the standard was within
0.1 airmass of the source observations. The standard star used was
BS8246 an AO dwarf whose spectrum only shows hydrogen absorption lines
in the observed wavelength range at this spectral resolution.

The initial data reduction of the spectrum (i.e bias, dark and flat
correction) was performed using the ORAC-DR\footnotemark[1] UIST
package. The resulting spectral image was then manually cleaned to
remove cosmic ray artifacts and bad pixels.  To produce a sky
subtracted spectrum the positive and negative nod positions were
extracted and subsequently subtracted from each other.  The region
known as N0 (cf. Fig.~\ref{morphology}) was extracted (0.80
arsec$^2$). The N1 region (the reflection nebulosity to the
north-west) proved too faint for any useful analysis. Wavelength
calibration was carried out using the argon arc lamp frames.

To correct the spectrum for atmospheric absorption, it was divided
through by that of the standard star. To reduce the impact of the
hydrogen recombination Brackett absorption lines on the final object
spectrum, any strong absorption lines were removed and replaced with
an interpolated continuum prior to division.  The interpolation of the
underlying continuum was determined via Gaussian or Lorentzian
(depending on the shape of the line) fitting to the lines.

We did not flux calibrate the data as the observations were taken
through thin cloud and conditions were not photometric. However, the
shape of the continuum should be reasonably well reproduced.

All of the data reduction steps described above were performed using
the STARLINK package FIGARO. Despite the fact that V645 Cyg is a
well-studied object, to our knowledge, the near-infrared spectra under
discussion, both in terms of wavelength coverage and resolution,
constitute the best data available for this object.

\subsection{Optical Spectroscopy}

We have obtained intermediate resolution optical spectra of V645 Cyg
from the Isaac Newton Group
archive\footnote{http://archive.ast.cam.ac.uk/ingarch/}. The
observations of H$\beta$ obtained on the 10th July 1993 and H$\alpha$
on the 19th December 1994 were taken with the Intermediate Dispersion
Spectrograph (IDS) on the 2.5 metre Isaac Newton Telescope. For the
H$\beta$ spectrum the 500mm camera, EEV5 1242x1152 element array and
the R632V grating was used. This gave a spectral resolution of 130
\rm{kms}$^{-1}$. A similar setup, but now with the Tek3 1024x1024
pixel array and R1200B grating was used for the H$\alpha$
observations, which gave a velocity resolution of 80 \rm{kms}$^{-1}$.

The data were cosmic ray, bias and flat corrected. Wavelength
calibration was provided by CuAr and CuNe arc frames.

\section{Results}
\label{results}

\subsection{Infrared Luminosity}
\label{rms}

The distance to V645 Cyg has been subject to considerable debate (see
\citealt{schulz} for a full discussion) with values ranging from 3.5
to 6 kpc. From our high sensitivity $^{13}$CO 2-1 line spectrum
\citep{ant} we determined the radial velocity of V645 Cyg to be
V$_{LSR}$$=$$-43.90$kms$^{-1}$,which is consistent with previous
studies (e.g \citealt{harvey}, \citealt{schulz}).  Combining this
value and a standard galactic rotation curve \citep{brand} we
calculate the kinematic distance to V645 Cyg to be 5.7 kpc
\citep{ant}. However, V645 Cyg is located within the second quadrant
($l=90-180\degr$) of our galaxy, where the non-circular motions in the
Perseus arm \citep{humphrey76} are significant.  Therefore, after
taking into account a proper motion of 10 \rm{kms}$^{-1}$ and the
uncertainty in the galactic rotation curve, we arrive at a distance of
5.7$\pm$1 kpc. This is on the higher end of the previous distance
estimates and consistent with earlier millimetre line data
(e.g. \citealt{harvey}). After including IRAS photometry, the total
infrared luminosity of V645 Cyg can be estimated to be between
2$-$7$\times 10^4$L$_{\odot}$. This corresponds to an early B-type for
the object, even for the lower distance of 3 kpc. Despite the
uncertainty in the distance, V645 Cyg is clearly an intrinsically
bright source.

\subsection{Near/Mid Infrared Imaging}

\begin{figure*}
\includegraphics[scale=0.45]{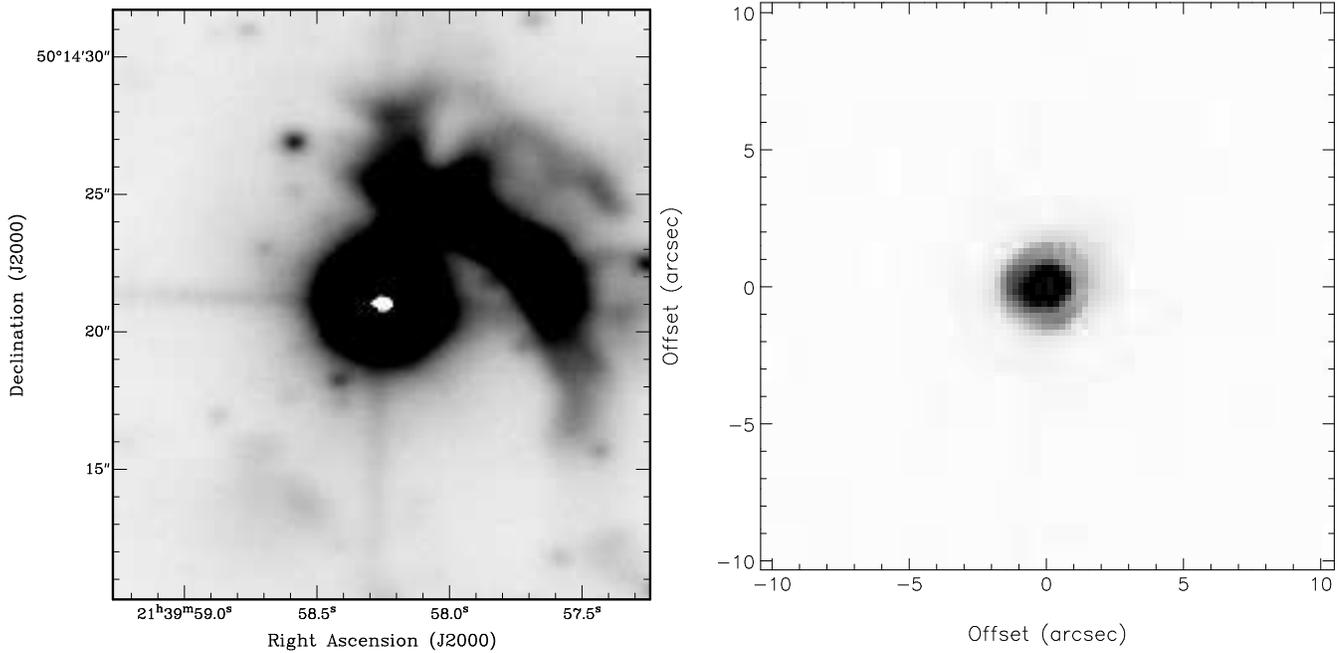}
\includegraphics[scale=0.55]{4839f1b.eps}
\caption{$K$ band UFTI image (left) and $N$ band Michelle image
(right) of V645 Cyg. In both images North is up and East to the
left. In the $K$ band image, V645 Cyg N0 is the central saturated
source and N1 is the extended reflection nebulosity to the the
north-west. In the $N$ band image the horizontal and vertical offsets
represent the R.A. and Dec offsets from the position of V645 Cyg
N0. The source may be slightly resolved in the north-west direction
compared to the seeing of approximately 1 arcsec. Notably no emission
is observed from component N1.}
\label{morphology}
\end{figure*}

In Figure \ref{morphology} we present a near-IR $K$ and mid-IR $N$
band image of V645 Cyg. The $K$ band image shows a bright point source
(named V645 Cyg N0) with a diffuse nebulosity (V645 Cyg N1)
approximately 5 arcsec to the north-west. The N0 component is a factor
of 200 brighter than the N1 component, and has a full width at half
maximum (FWHM) of approximately 0.48 arcsec. This is comparable to the
seeing and N0 can be considered a point source. The $N$ band image
shows mid-IR emission {\it only} from the N0 component, and no
emission from N1. A comparison with unresolved (i.e showing an airy
disk) objects observed on the same night, indicates the mid-IR
emission from V645 Cyg may be slightly resolved in the north-west
direction. This emission would most likely arise from hot dust
surrounding V645 N0, the other possibility is an ultracompact H {\sc
ii} region but the relative weakness of the radio continuum emission
does not support this notion.

The interpretation of previous optical imaging observations
(\citealt{cohen}; \citealt{goodrich}; \citealt{hamann}) was that N0
and N1 are knots of reflection emission from the star which is more
deeply embedded within the molecular cloud. However, our infrared
imaging observations show the N0 component to be a point source in the
near-IR. To within the errors of less than 2 arcsec, the sole source
of mid-IR emission is associated with N0 as well. In addition, the
slightly lower spatial resolution polarimetric measurements by
\cite{minchin} show a centro-symmetric polarization pattern pointing
toward a source close to N0 being the central source.  Given this
evidence, it seems clear that the N0 component is the star itself and
not simply a knot of reflection nebulosity like N1.

\subsection{Near-IR spectroscopy}
\label{uist}

\begin{figure*}
\centering
\includegraphics[scale=1.25]{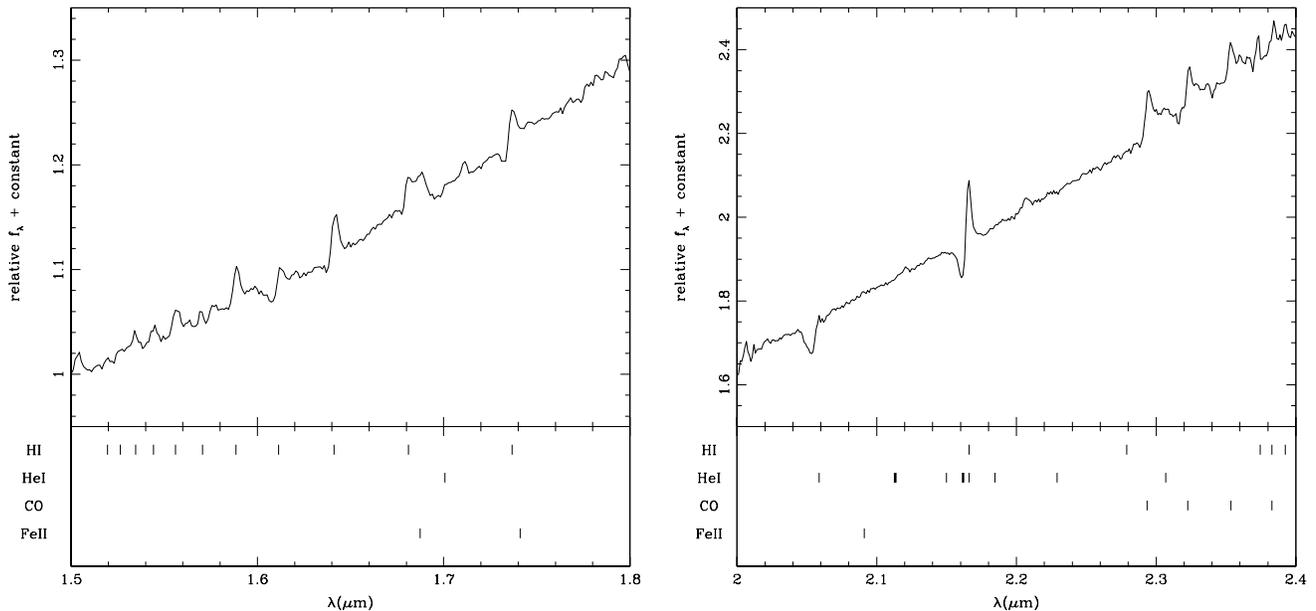}
\caption{$H$ (left) and $K$ (right) band region of V645 Cyg
spectrum. The wavelengths of common elements are indicated by vertical
lines in the lower panel. The spectrum is trimmed to 2.4$\mu$m, due to
strong atmospheric absorption beyond this wavelength. }
\label{spec}
\end{figure*}

In Figure \ref{spec} we present the final $H+K$ spectrum of V645 Cyg
N0. The spectrum is red, and displays a number of emission and
absorption lines due to H {\sc i}, He {\sc i}, CO, Fe  {\sc ii}, [Fe  {\sc
ii}] and Na {\sc i}. Perhaps most notable of these are the strong CO
bandhead emission at 2.3 $\mu$m, the P-Cygni profiles of the hydrogen
recombination lines and the strong blueshifted 2.058$\mu$m He {\sc i}
absorption line. We discuss the properties of the spectrum in more
detail below.

\subsubsection{Hydrogen}

As can be seen in Fig.~\ref{spec}, the Brackett series hydrogen
recombination lines show P-Cygni profiles. Such line profiles have
previously been observed in optical hydrogen lines (\citealt{cohen};
\citealt{hamann}; \citealt{goodrich}), however the velocities implied
are nearly double that previously observed.  The blue absorption wing
extends to velocities of nearly -2000 \rm{kms}$^{-1}$. This may imply
an increase in the wind velocity or it could simply be an optical
depth effect.

\begin{figure*}
\centering \includegraphics[scale=0.85]{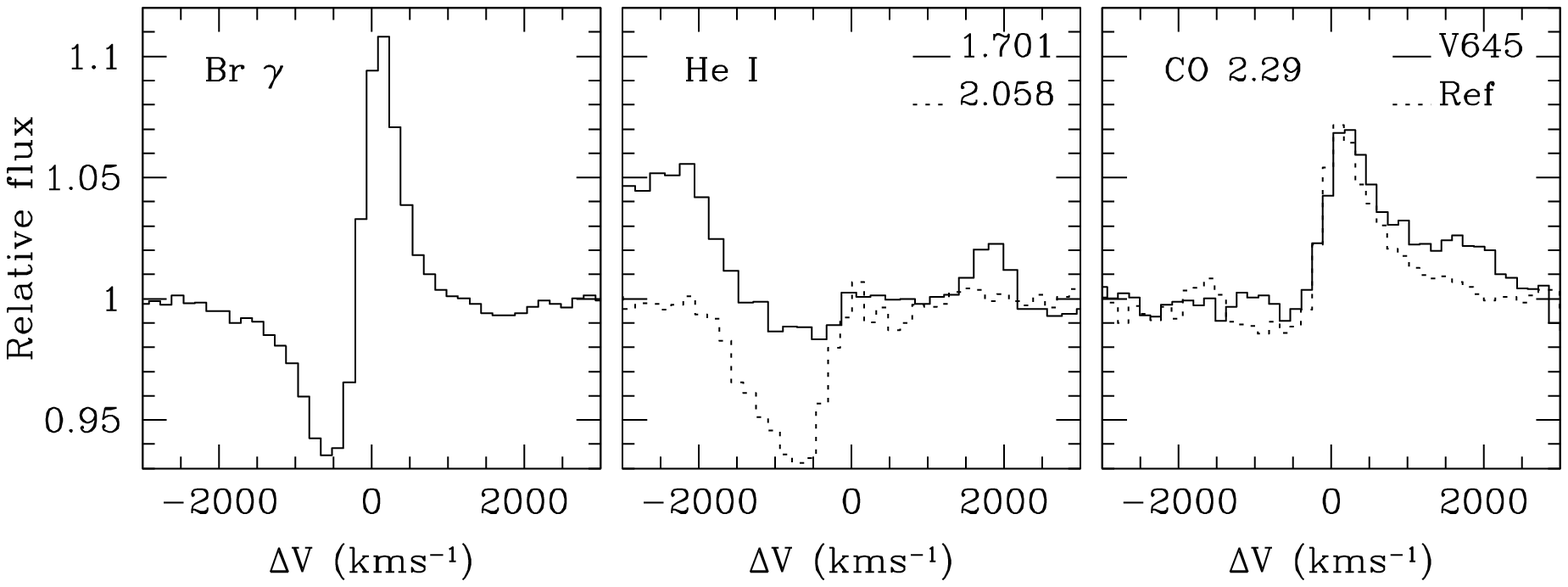}
\caption{The continuum normalised profiles of Brackett $\gamma$
(left), He {\sc i} (centre), $^{12}$CO 2.29 (right).  In the left
panel, Br$\gamma$ shows a P-Cygni line profile, with absorption up to
2000 \rm{kms}$^{-1}$. In the centre panel, the 1.701 He {\sc i} line
is indicated by the solid line and the 2.058 line by the broken
line. Both He {\sc i} lines show blue shifted absorption with a
similar velocity to each other, however the 1.701 line also shows a
red-shifted emission component. The feature at -2000\rm{kms}$^{-1}$ in
the 1.701 He {\sc i} line is the wing of the 1.680 Fe {\sc ii}
line. In the right panel, the $^{12}$CO bandhead profile of V645 Cyg
is represented by a solid line. As a guide to the normal CO bandhead
profile, a scaled CO emission profile of G195.6495-00.1057 a candidate
MYSO from the RMS sample is indicated by the broken line. A clear
emission feature is observed at +1800 \rm{kms}$^{-1}$.}
\label{profiles}
\end{figure*}

\begin{figure*}
\centering 
\includegraphics[scale=0.75]{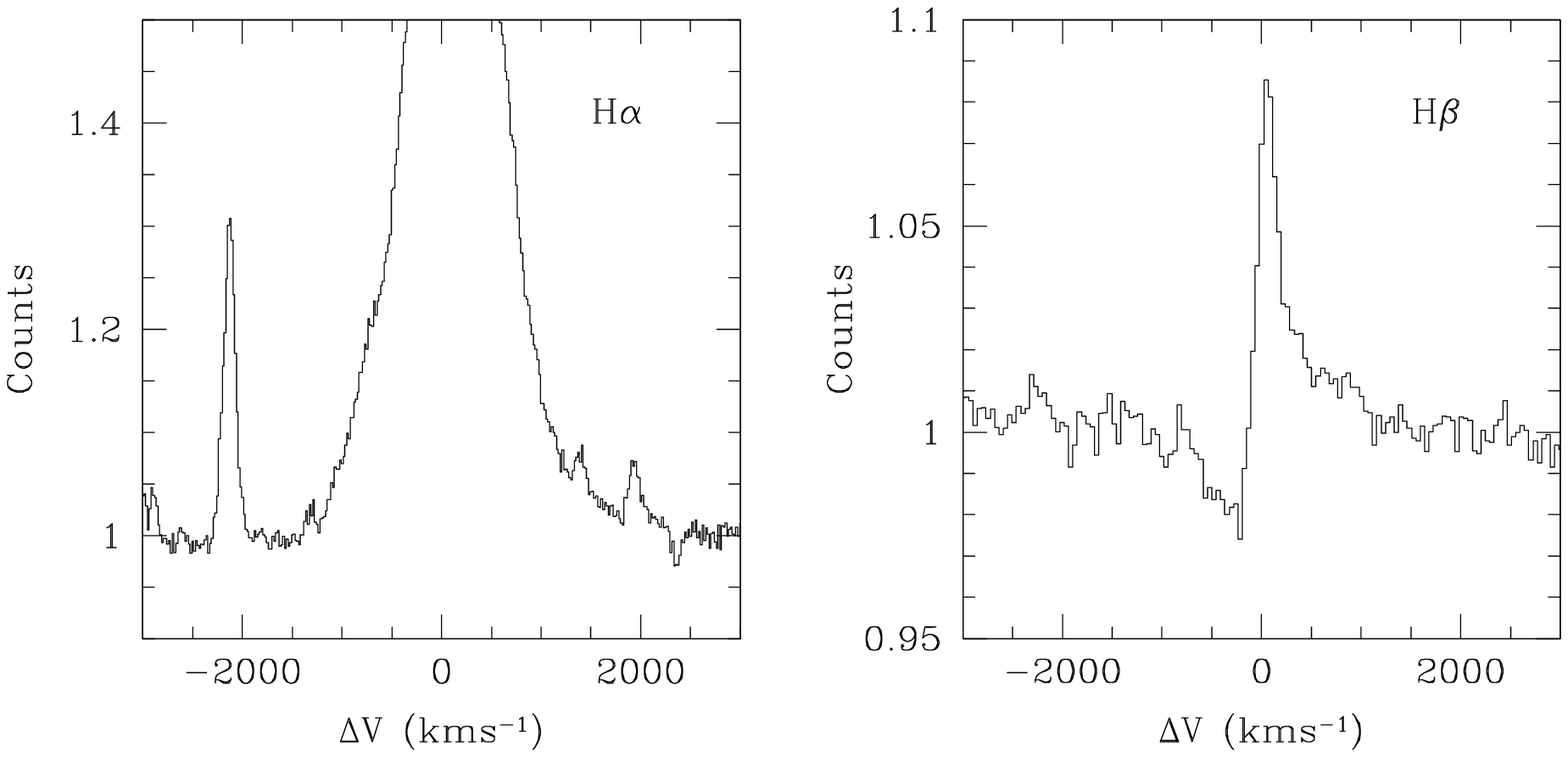}
\caption{The continuum normalised profiles of H$\alpha$ (left) and
H$\beta$ (right). The H$\alpha$ line shows a red-shifted component
with a velocity of +1800\rm{kms}$^{-1}$, whilst no such component to
the H$\beta$ line is present. The strong emission line at
-2250\rm{kms}$^{-1}$ in the H$\alpha$ spectrum is the 6516 Fe {\sc ii}
line and the weak emission feature blended with the red wing of
H$\alpha$ is 6584 [N {\sc ii}].}
\label{opticalspec}
\end{figure*}

\subsubsection{Helium}

The most obvious He {\sc i} line in the spectrum is the strongly
blue-shifted (again up to -2000 \rm{kms}$^{-1}$) absorption line at
2.058 $\mu$m (plotted together with the 1.70 $\mu$m line in
Fig.~\ref{profiles}). The absence of an emission component makes this
line notably dissimilar to the P-Cygni profile lines in the spectrum,
and is indicative of a much smaller line forming region than the
hydrogen lines. The propensity of this He {\sc i} line to manifest
itself in absorption is well known \citep{drew}, and is attributable
to the metastable character of its lower level.

The other He {\sc i} line in the spectrum, at 1.70 $\mu$m, has weak
blueshifted absorption, but no obvious emission at the systemic
velocity. The emission that is present at the bluest velocities is due
to Fe {\sc ii}, not the He {\sc i} line itself. As the line is
optically thin, it is much weaker in absorption than the 2.058 $\mu$m
line. The emission feature at $\sim$1800 kms$^{-1}$ can not be
identified with any known lines listed, instead, we will argue that
this is red-shifted helium emission.  Using the presence and absence
of helium lines in the spectrum, we can spectrally type the object.
To this end we compare with the spectral atlas of \cite{hanson}. The
mere presence of He {\sc i} absorption argues for an early spectral
type, but the complete absence of the He {\sc ii} 2.19 $\mu$m puts it
towards the later end of the O star spectrum. The absorption of He
{\sc i} 1.70 and 2.058 $\mu$m lines and weak emission from He {\sc i}
2.11 $\mu$m gives the best match for a late O ($\sim$8.5) supergiant
type object. Because the lines are most likely to be formed in the
wind, it is not the spectral type of the star itself that we have
determined, rather it is more plausible that the wind is optically
thick and that the extended, dense, wind creates a pseudo-photosphere
with this early type spectrum. This solves the potential puzzle why
the outflow velocities are much higher than might be expected from a
supergiant, as the embedded object where the flow finds its origin is
much smaller. It also implies that we deal with a much hotter object
than implied by the spectral type itself.

\subsubsection{Variable CO Bandhead Emission}

The spectrum shows strong first overtone $^{12}$CO bandhead emission,
the CO 2-0, 3-1, 4-2 and possibly even 5-3 bands are visible in the
spectrum.  No evidence for $^{13}$CO emission is found. CO emission
was first detected, with an equivalent width of 3.8$\AA$, by
\cite{geballe} in 1985 but the subsequent searches of \cite{carr} in
1986 and \cite{biscaya} in 1997, determined strict 3$\sigma$ upper
limits to the equivalent width of 1.3$\AA$ and 0.79$\AA$
respectively. This together with our data (equivalent width 5.2$\AA$)
obtained in 2003, indicates that the object has a strong variability
in its CO emission. Unfortunately the infrequency of these
observations makes it difficult to make any conclusions about the
timescale of this variability. However, from the 1985 detection and
subsequent 1986 non-detection we can say that the CO emission must
switch off relatively quickly.

The CO bandhead emission also shows an unusual profile.  Figure
\ref{profiles} zooms in on the CO 2-0 bandhead, and compares this with
a ``typical'' emission line from another of our target stars (Clarke
et al. in preparation). The lines appear to have a double peak, the
weaker red-shifted peak having a velocity of 1800\rm{kms}$^{-1}$.
This may be related to the similarly redshifted He~{\sc i} emission,
but would rule out rotation as the single cause of the kinematics.

\subsubsection{Other metals}

Fluorescent Fe {\sc ii} emission is present at 1.680 $\mu$m while
forbidden [Fe {\sc ii}] emission at 1.643 $\mu$m is also present. The
latter is blended however with the Brackett 12-4 line. The emission at
1.711 $\mu$m seems unlikely to be associated with the [Fe {\sc ii}]
line at this approximate wavelength as its intensity should be less
than a tenth of the 1.644 $\mu$m line. This emission is therefore more
likely the red-shifted component of the nearby 1.701$\mu$m He {\sc i}
emission, of which its velocity agrees well. A weak Na {\sc i} emission
line can be also seen at 2.208 $\mu$m. The Fe {\sc ii}, [Fe {\sc ii}]
and Na {\sc i} line profiles all appear unresolved and certainly do not
show the same velocity structure as the hydrogen or helium lines. This
finding complements the optical studies of \cite{hamann} which also
found no structure in the iron lines.

\subsubsection{High Velocity Red-Shifted emission}

As already alluded to above, a red-shifted emission component is
present in the 1.701$\mu$m He {\sc i} line and in the CO bandhead
emission lines. The velocity of this component is approximately 1800
\rm{kms}$^{-1}$.  Br$\gamma$ does not display the feature and this
would be in apparent contradiction with the fact that we can see it in
both lower and higher excitation lines.  However, the component may
have been observed previously in optical hydrogen lines by
\cite{hamann}. They specifically highlight an unidentified emission
feature redshifted by 1800 \rm{kms}$^{-1}$ from the H$\alpha$ line in
their optical spectrum of the object. This is in agreement with the
features that we find in helium and carbon-monoxide.

In Figure \ref{opticalspec} we present H$\alpha$ and $H\beta$ spectra
of V645 N0. The H$\alpha$ emission line is notably broader than the
H$\beta$ line, and is attributable to the H$\alpha$ being optically
thick, enhancing the appearance of the electron scattered wings. In
this particular spectrum the H$\alpha$ line also possesses the
red-shifted counterpart with a velocity of approximately 1800
\rm{kms}$^{-1}$, this agrees well with the red features in the near-IR
CO and He {\sc i} lines.  The equivalent width of the red-shifted
H$\alpha$ component is 2\% of the main line.  The spectrum of $H\beta$
does not show any evidence for a red-shifted counterpart. The
signal-to-noise at H$\beta$ is less than at the redder wavelengths,
but this should not necessarily explain the absence of the red-shifted
feature. However, as it is not unreasonable to assume that the
extinction towards the feature is much higher at the bluer
wavelengths, relative strength of the receding component compared to
the ``main'' feature could be much less at H$\beta$ than at H$\alpha$,
explaining its non-detection. Based on a similar argument we might
expect the detection of this feature at Br$\gamma$ where it is most
certainly not seen. Pending proper radiative transfer computations
this will be difficult to predict, but we note that H$\alpha$ is an
optically thicker line than Br$\gamma$, enhancing the relative
strength of any other, but optically thin, feature from the same line.

\section{Discussion and Conclusions}
\label{discuss}

We have presented new multi-wavelength observations of V645 Cyg.  Our mid/near
infrared imaging data imply that the stellar N0 component of V645 Cyg is the
location of the mid infrared emission from this source.  N0 is also the
exciting star of the diffuse reflection nebulae V645 Cyg N1, as previously
shown by \cite{minchin}. 

The P-Cygni like absorption of the Brackett hydrogen lines is
consistent with a powerful stellar wind.  The velocity of this wind is
typical of that expected from a young OB type star but stands out for
a massive young stellar object, where much lower outflow velocities
are the norm (\cite{drew}: many massive young stellar objects show no
P-Cygni features at all). Indeed, when taking the crude spectral
classification of an O8.5 supergiant at face value, the spectral class
itself would be consistent with low outflow velocities, as typically,
supergiants have lower escape velocities than dwarf stars. The fact
that the classification is based on absorption lines extending to 2000
kms$^{-1}$ strongly suggests, however, we are dealing with a rapidly
expanding pseudo-photosphere.  This notion is corroborated by the
luminosity estimate for V645 Cyg, which indicates it to be a high mass
object but not a supergiant. For the accepted distances (roughly 3-6
kpc), the luminosity falls within the range 2$-$7$\times
10^4$L$_{\odot}$. This is consistent with early B/late O star
zero-age-main-sequence bolometric luminosities, but not with a
supergiant luminosity.  The fact that the high excitation He {\sc i}
lines do not show visible emission indicates that the acceleration
region is very small, and that the terminal velocities are reached
quickly.  These high velocities are not often observed in MYSOs. Low
outflow velocities in MYSOs can be explained by radiatively driven
disk-winds (cf. \citealt{drewproga}). Were V645 Cyg to belong to the
MYSO class, it would would appear that the circumstellar disk has
disappeared and the wind is a purely radiatively driven stellar
wind. It may thus well be a transitionary object between the MYSO
phase and the ensuing H {\sc ii} region phase.  In this case the wind
must still be thick enough to absorb the ionising radiation from the
central star since there is as yet no evidence for an H {\sc ii}
region being present, and a star of this luminosity should be able to
generate such a region (unlike lower mass Herbig type stars).  The
relatively low extinction towards the source is consistent with this
picture of an `evolved' MYSO.

Intriguingly, the spectrum of V645 Cyg shows red-shifted emission
components to the He {\sc i}, hydrogen recombination and CO lines. The
velocity of this redshifted component is remarkably similar to the
blueshifted velocities measured in the P Cygni absorption troughs. We
therefore conclude that the redshifted features correspond to the
hitherto unidentified receding part of the flow.  The relative
weakness of the emission could be explained by obscuration due to a
dense disk or simply dust extinction within the nebula.

The observation of variable CO bandhead emission is not particularly
unusual, as \cite{biscaya} found many YSO's show CO bandhead
variability on a range of timescales. The variability of CO bandhead
emission is related to the special conditions that are required to
produce it. The warm, dense and neutral material that is required to
produce CO bandhead emission is generally only found in dense rotating
disks or mildly shocked regions.  The observation of a high velocity
red-shifted component to the CO emission would appear to imply that
its emission is related to the stellar wind of the star rather than a
rotating disk. This introduces another puzzle, as an 1800\rm{kms}$^{-1}$
flow shocking the interstellar medium would most likely destroy any
molecular CO.  It may be possible that we are witnessing the complex
transition zone between the stellar wind and the larger scale
molecular outflow.  In this case the CO emission should be tied
strongly to the strength of the stellar wind, and the existing data do
suggest this may be true.

V645 Cyg therefore presents a variety of problems as regards its exact
nature.  Further observations of the clearly variable wind would be
helpful in illuminating some of these.  Overall it may represent a
relatively rare class of transition objects between a genuinely
massive young stellar object and a normal young Oe type star in a weak
H{\sc ii} region.

\begin{acknowledgements}

We would like to thank the referee, Rafael Bachiller, for his helpful
comments. UKIRT is operated by the Joint Astronomy Centre on behalf of
the UK Particle Physics and Astronomy Research Council (PPARC). The
James Clerk Maxwell Telescope is operated by the joint Astronomy
Centre on behalf of the Particle Physics and Astronomy Research
Council of the United Kingdom, the Netherlands Organisation for
Scientific Research and the National Research Council of Canada. The
Issac Newton Telescope is operated on the island of La Palma by the
Isaac Newton Group in the Spanish Observatorio del Roque de los
Muchachos of the Institutio de Astrofisica de Canarias. This
publication makes use of data products from the Two Micron All Sky
Survey, which is a joint project of the University of Massachusetts
and the Infrared Processing and Analysis Center/California Institute
of Technology, funded by the National Aeronautics and Space
Administration and the National Science Foundation.

\end{acknowledgements}

\end{document}